# BioEnvSense: A Human-Centred Security Framework for Preventing Behaviour-Driven Cyber Incidents

Duy Anh Ta[1], Farnaz Farid*[2], Farhad Ahamed[2], Ala Al-Areqi[2], Robert Beutel[2], Tamara Watson[2], Alana Maurushat[1,2]

[1] *School of Computer, Data and Mathematical Sciences*
[2] *School of Social Sciences*
*Western Sydney University*
Farnaz.Farid@westernsydney.edu.au

**Abstract**

Modern organizations increasingly face cybersecurity incidents driven by human behaviour rather than technical failures. To address this, we propose a conceptual security framework that integrates a hybrid Convolutional Neural Network–Long Short-Term Memory (CNN–LSTM) model to analyze biometric and environmental data for context-aware security decisions. The CNN extracts spatial patterns from sensor data, while the LSTM captures temporal dynamics associated with human error susceptibility. The model achieves 84% accuracy, demonstrating its ability to reliably detect conditions that lead to elevated human-centred cyber risk. By enabling continuous monitoring and adaptive safeguards, the framework supports proactive interventions that reduce the likelihood of human-driven cyber incidents.

## 1 Introduction

Cybersecurity incidents continue to increase globally, and human behaviour has emerged as a primary contributing factor. Research indicates that approximately 95% of cybersecurity breaches are linked to human error or risky user actions (Verizon 2023). These incidents are not limited to malicious intent but also include everyday mistakes influenced by stress, distraction, or an unstable mental state. As everyone has become more digitally interconnected than ever, both companies and individuals are facing heightened exposure to security vulnerabilities (Sasse and Flechais 2019).

Particularly, insider threats, which are risks posed by individuals within an organization, whether intentionally or unintentionally, account for a significant proportion of security incidents and represent a critical concern. Cybersecurity Insiders reports that 83% of organizations experienced at least one insider attack in 2024, many of which were unintentional rather than deliberate attacks (Cybersecurity Insiders 2024). Additionally, Insider Risk Research reports that the average cost per insider threat incident has risen to 1.034 million AUD, with the annual average cost reaching around 26.61 million AUD (Ponemon Institute 2024). Existing mitigation measures for such unintentional actions, such as security training, awareness campaigns, and automated reminders, provide only partial protection but fail to account for users' real-time cognitive or emotional state (Greitzer et al. 2014). Likewise, biometric authentication methods such as facial recognition and fingerprint scanning primarily focus on identity verification rather than the behavioural or emotional factors that influence secure system use (Jain et al. 2016). This gap highlights the need for systems that are responsive to users and adaptive to people, for example, a security framework that can guide users in a timely fashion based on their stress and environment, leading to safe cyber interactions. There is emerging evidence that soft biometrics, such as heart rate variability, micro-expressions, blood pressure, and galvanic skin response, in conjunction with contextual environmental information such as noise, lighting, or temperature, can indicate cognitive workload and emotional state (Healey and Picard, 2005; Roda and Thomas, 2006). Including these indicators in security frameworks could increase organizations' ability to proactively identify times when users are more likely to commit harmful errors.

To address this issue, this work, an extension of our previous work (Farid, 2023), proposes a conceptual security framework that integrates soft biometric data with environmental sensing to estimate users' emotional and cognitive states in real time. When elevated risk is detected, the framework can initiate protective actions, such as issuing warnings, providing additional instructions, enabling backup support, or temporarily restricting access. This framework aims to reduce non-intentional insider threats while maintaining usability. To this end, the paper is structured as follows: Section 2 illustrates the methodology that informs this work; Section 3 discusses model development and evaluation; Section 4 details the model deployment; Section 5 discusses limitations and future directions of this work; Section 6 provides a literature review; and Section 7 finally concludes the paper.

## 2 Methodology

This section discusses the methodology that has been used in this research. We employ a five-phase framework that includes data processing, data labelling, model development, model evaluation and model deployment.

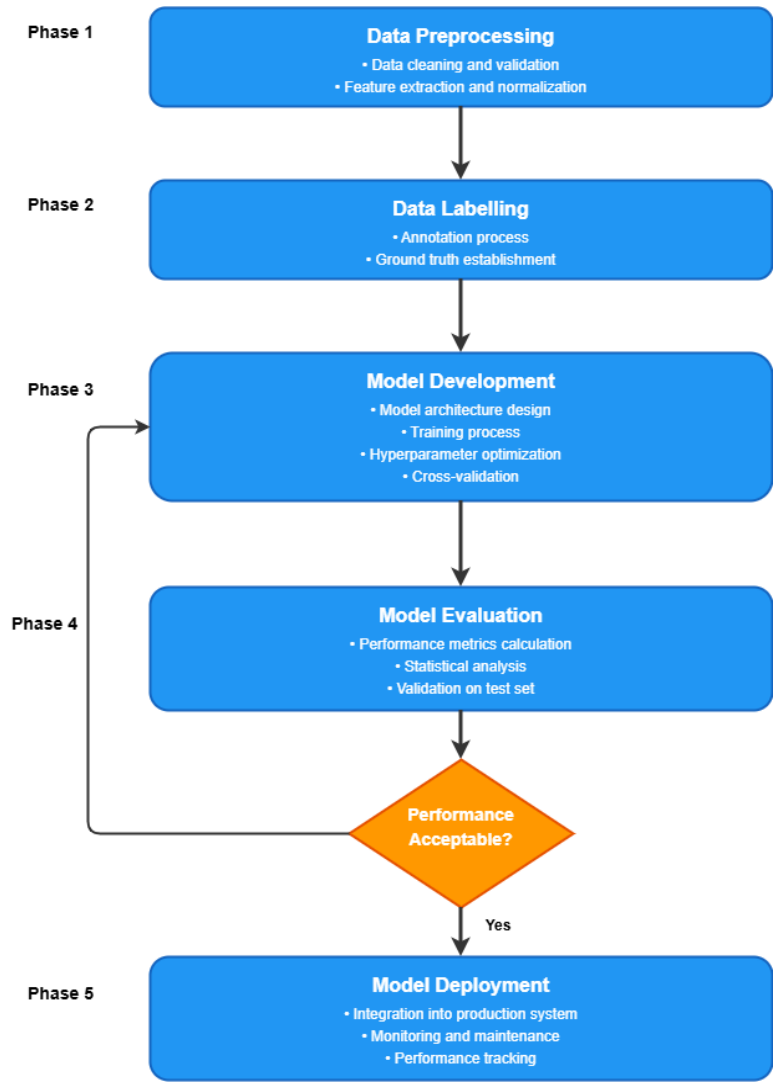

*Fig. 1* Methodology Framework

## 2.1 Data Collection and Data Preprocessing

This study uses data from Anicai and Shakir (2023), which is publicly available for academic and industry research. The dataset provides a comprehensive view of how the human body reacts to various environments, based on an experiment involving 14 participants over 600 minutes. Thus, this repository was used as the best one for the experiment as it captures simultaneous markers of physiology (Heart Rate, GSR) as well as of environment (Light, Noise, Temperature), something that is essential for the current study to be able to analyze how external conditions and the internal stress responses are interacting with each other.

**Table 1:** Dataset specifications

| Metric | Specification |
| --- | --- |
| Total Participants | 14 |
| Sampling Rate | 1 Hz |
| Total Duration | ~600 minutes (approx. 43 min per subject) |
| Sensors Used | Heart Rate (BPM), GSR (µS), Temperature (°C), Light (Lux), Sound (dB) |
| Input Shape | 5 Features (Multivariate Time-Series) |
| Window Size | 30 Time-Steps (30 seconds) |

To ensure data quality and compatibility with the model, raw sensor streams underwent a rigorous five-stage preprocessing pipeline before model ingestion:

1. Data Cleaning: Corrupted records and rows containing null values were removed via list-wise deletion to ensure temporal continuity.
2. Feature Selection: The dataset was reduced to five core predictors: Heart Rate, GSR, Temperature, Lux, and Sound.
3. Z-Score Standardization: Data was normalized per subject to mitigate baseline physiological variability and centre signals around zero.
4. Sliding Window Segmentation: Continuous streams were segmented into 30-second overlapping windows with a 1-second stride to capture temporal stress trends.
5. Subject-Independent Splitting: Participants were partitioned into distinct training (80%) and validation (20%) sets to test generalization and prevent data leakage rigorously.

**2.2 Label Definition**

As the original dataset lacked cognitive or emotional labels, a custom labelling scheme was developed to classify user states based on physiological and environmental inputs. The labelling framework integrates two major components: cognitive performance, derived from environmental conditions, and physiological stress, calculated from Heart Rate (HR) and Galvanic Skin Response (GSR).

For cognitive performance, environmental variables including temperature, noise levels, and illuminance were mapped to cognitive performance scores on the scale from 0 to 1 based on scientific evidence. Here are the descriptions and range for each input:

- Temperature: Optimal cognitive performance is usually found around 20-24°C for office tasks; performance declines at higher temperatures (≥25°C) and when too cold (Seppänen et al. 2006; Lan et al. 2011). The top range is set to 20–24°C, with performance configured to drop toward 0 below ~16°C and above ~30°C.

**Table 2:** Weighting Factors for Temperature

| Temperature | Score |
|---|---|
| <18°C | 0.3 |
| 18–20°C | 0.6 |
| 20–24°C | 1 |
| 24–26°C | 0.8 |
| 26–28°C | 0.5 |
| >28°C | 0.2 |

- Sound: Quiet environments (< ~40-45 dB) support best performance; as noise rises (speech, babble, traffic), working memory, attention and reaction time deteriorate, larger effects above ~55 dB and especially at high levels (Szalma and Hancock 2011; Shield and Dockrell 2008). Best performance occurs at ≤40 dB with progressive decline to 0 at very high noise (≥85 dB).

**Table 3:** Weighting Factors for Sound

| Sound (dB) | Score |
|---|---|
| 0–30 | 1 |
| 30–50 | 0.9 |
| 50–60 | 0.7 |
| 60–75 | 0.4 |
| >75 | 0.2 |

- Light (for outdoor lighting): Outdoor illuminance can range from 10,000 lux (in shade) to >100,000 lux (in direct sunlight). The relationship with cognition is non-linear, some daylight exposure improves alertness and mood, but too much brightness or glare causes eye strain and mental fatigue (Boubekri et al. 2014; Figueiro et al. 2017).

**Table 4:** Weighting Factors for Light

| Lux | Score |
|---|---|
| 0–500 | 0.2 |
| 500–2000 | 0.4 |
| 2000–10000 | 1 |
| 10000–20000 | 0.7 |
| >20000 | 0.3 |

The cognitive score was computed as the average of the three environmental scores.

For physiological stress, it is estimated using normalized HR and GSR, both of which serve as indicators of sympathetic nervous system activation (Healey and Picard 2005). The reason for using normalized data but not the normal range for most people is that each dataset is particular to one person with unique characteristics. Therefore, using the normal range could lead to incorrect labelling. HR and GSR were independently normalized to a 0-1 scale, where higher values indicate increased arousal. Based on evidence from the original dataset's classification, showing that stress has a stronger influence on error likelihood than cognitive load, HR was weighted at 70% and GSR at 30% in computing the stress score (Anicai and Shakir 2023).

The overall risk score was computed using a weighted combination of stress and cognitive components:

$$Overall\ score = 0.7 \times Stress\ Score + 0.3 \times (1 - Cognitive\ Score) \quad (1)$$

This formulation reflects the stronger role of physiological stress in promoting human error, supported by previous findings showing that stress reduces error awareness and increases preference for less demanding behaviours (Staal 2004; Hockey 1997).

**Table 5** Final label assignment

| Overall Score | Label | Meaning |
|---|---|---|
| 0.00–0.25 | Run as usual | Low risk |
| 0.25–0.50 | Show warning | Moderate risk |
| 0.50–0.75 | Limit access | High risk |
| 0.75–1.00 | Inform backup person | Critical risk |

## 3 Model Development and Evaluation

To identify the optimal architecture for risk detection, four distinct modelling strategies were evaluated, each yielding unique performance outcomes and generalization limitations. The initial phase investigated ensemble stacking techniques: the first architecture (Stack 1) utilized a combination of Random Forest, Gradient Boosting, and Logistic Regression as base learners (Layer 0), aggregated by a Logistic Regression meta-learner (Layer 1) (Wolpert 1992). The second architecture (Stack 2) enhanced this ensemble approach by incorporating a Neural Network at Layer 0 alongside Random Forest and Gradient Boosting to capture non-linear complexities better. Transitioning to deep learning methodologies for temporal sequence analysis, the third and fourth models employed a hybrid deep learning framework that integrated CNNs for spatial feature extraction with LSTM layers for temporal reasoning (Hochreiter and Schmidhuber 1997), with one of them exhibiting data leakage. Comparative analysis of these architectures revealed significant trade-offs between computational complexity and subject-independent accuracy, ultimately guiding the selection of the final model.

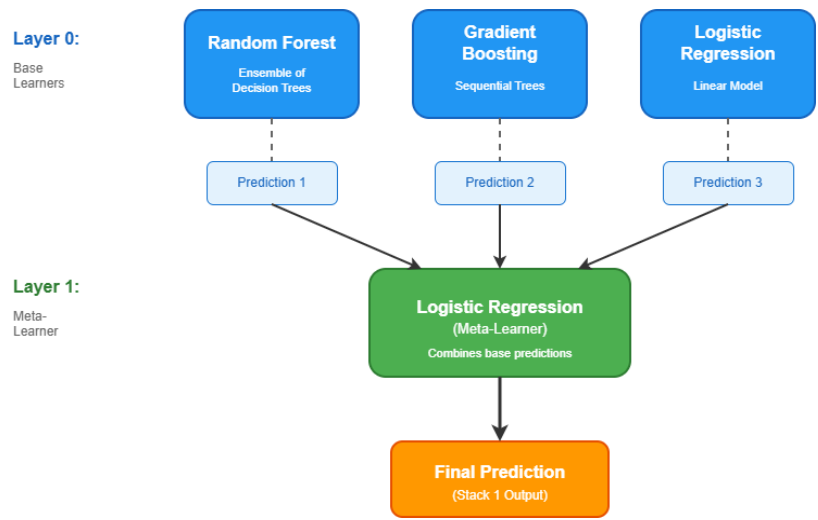

**Fig. 2** Stack 1 Ensemble Architecture

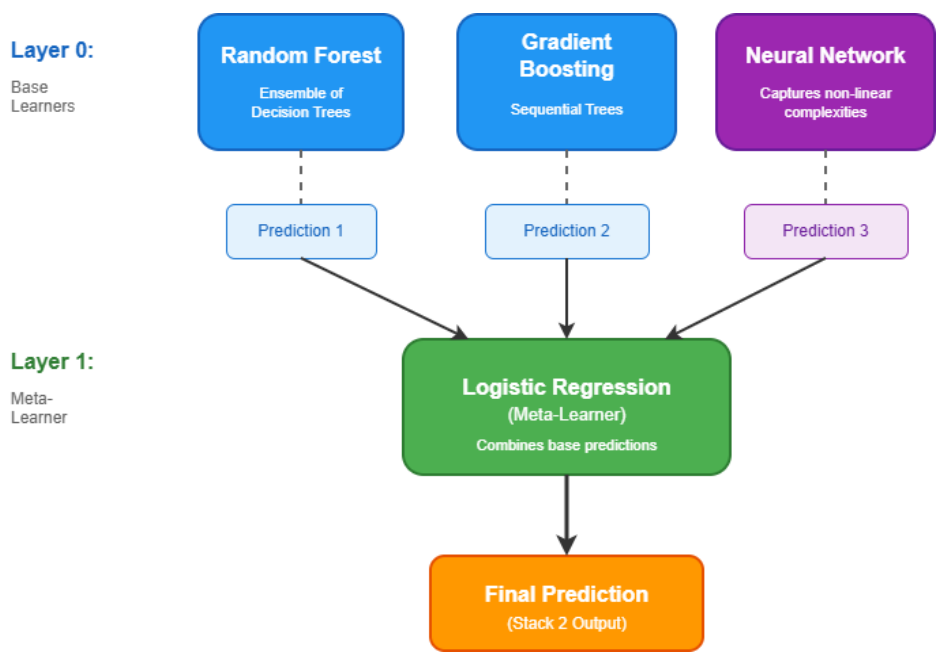

**Fig. 3** Stack 2 Ensemble Architecture

Evaluation of the ensemble stacking architectures (model 1 & model 2) initially yielded exceptionally high hold-out test accuracies of 97-98%. However, further diagnostic analysis using 5-fold cross-validation revealed a sharp drop in performance, with mean accuracies dropping to approximately 68%. This significant discrepancy indicates that the high initial scores were artifacts of data leakage inherent to random splitting in subject-dependent contexts, rather than true generalization (Kapoor and Narayanan 2023). Learning curve analysis further confirmed this diagnosis; both architectures exhibited a distinct high-bias convergence pattern, where training scores plummeted to meet validation scores at a suboptimal 70% threshold. This plateau suggests that the static, non-sequential feature representations used in these traditional ensembles imposed a hard performance ceiling. Consequently, despite their statistical complexity, these models proved unsuitable for real-world applications, as they lacked the temporal reasoning needed to distinguish dynamic stress events from benign physiological variations across individuals.

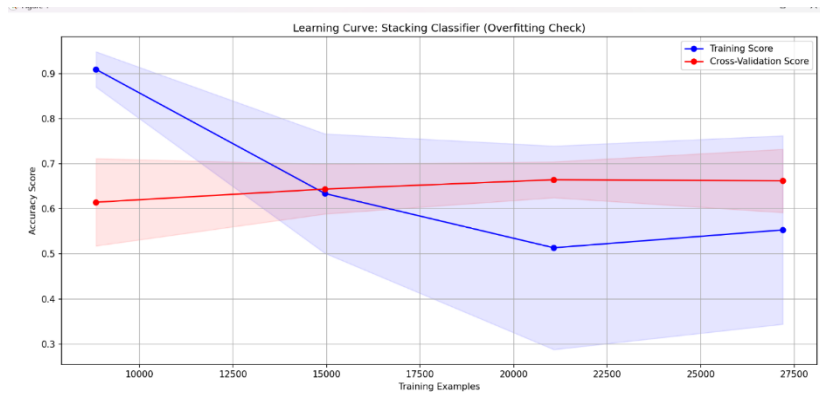

**Fig. 4** Learning curve of model 1

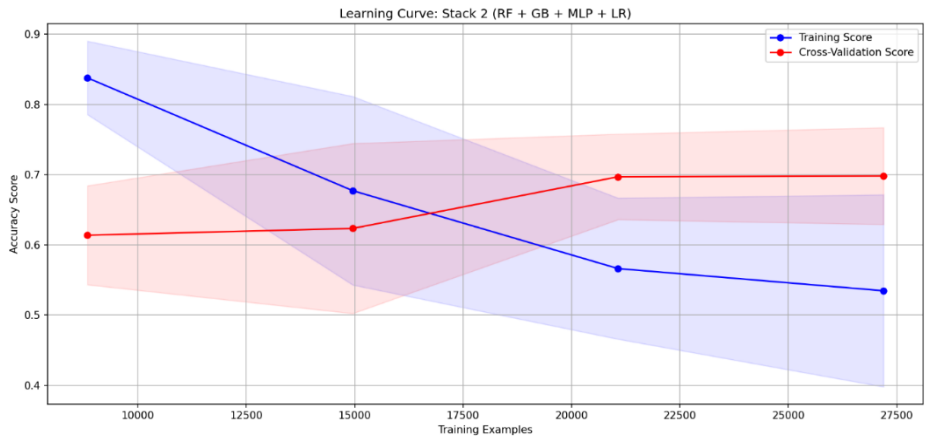

**Fig. 5** Learning curve of model 2

To address the inherent limitations of static ensemble classifiers, which treat physiological measurements as isolated snapshots, a Hybrid Deep Learning architecture (CNN-LSTM) was implemented. This selection was driven by the operational nature of the proposed application, which continuously captures physiological and environmental data at sequential time steps. Unlike random forests, which cannot perceive the order of events, the CNN-LSTM architecture is designed to model stress as a dynamic evolution rather than a static state (Sharma et al. 2016; Chao et al. 2019). The Convolutional Neural Network (CNN) layers are employed first to extract local morphological features (such as sudden spikes in GSR or Heart Rate) from the raw sensor windows, while the Long Short-Term Memory (LSTM) layers subsequently analyze these features over time to capture long-term dependencies (LeCun et al. 2015). This combination allows the system to distinguish between momentary physiological arousal and genuine, sustained stress events.

Our initial experimentation with Model 3 employed a subject-dependent approach that faced data leakage through inappropriate randomization of temporal windows prior to train-test splitting. This methodology allowed physiological windows from identical subjects, often recorded seconds apart, to appear in both training and testing partitions, enabling the model to exploit subject-specific physiological signatures rather than learning generalizable stress indicators. The resulting near-perfect synchronization between training and validation loss curves confirmed memorization rather than legitimate pattern recognition.

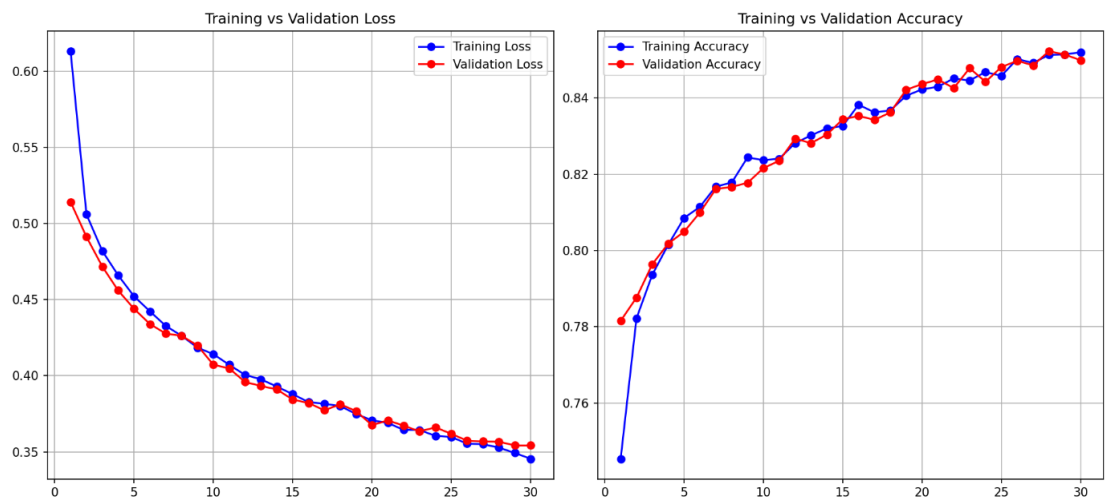

**Fig. 6** Loss and accuracy of training and validation of model 3

To address this fundamental flaw, Model 4 implements a rigorous subject-independent framework with participant-level data partitioning, ensuring complete separation between training and testing subjects (Schmidt et al. 2018). Specifically, the model processes sequential sensor data (window size t=30s) through a specific pipeline designed to handle noise and temporal dependencies:

- Feature Extraction: A 1D Convolutional layer (64 filters, kernel size=3) first extracts local morphological features, effectively filtering high-frequency noise from the physiological sensors.
- Temporal Processing: A subsequent LSTM layer (128 units) models the time-dependent progression of stress markers.
- Regularization: To mitigate overfitting, the architecture employs L2 regularization ($\lambda=0.01$) on kernels and Dropout layers (0.1 post-CNN, 0.3 post-LSTM).

The validation accuracy trends upwards, peaking at approximately 84%. The fluctuations observed in the validation metric are characteristic of the high dropout rate (0.3), which forces the network to learn robust features by randomly deactivating neurons during training. Crucially, the validation loss does not diverge significantly from the training loss, indicating that the L2 regularisation successfully prevented overfitting.

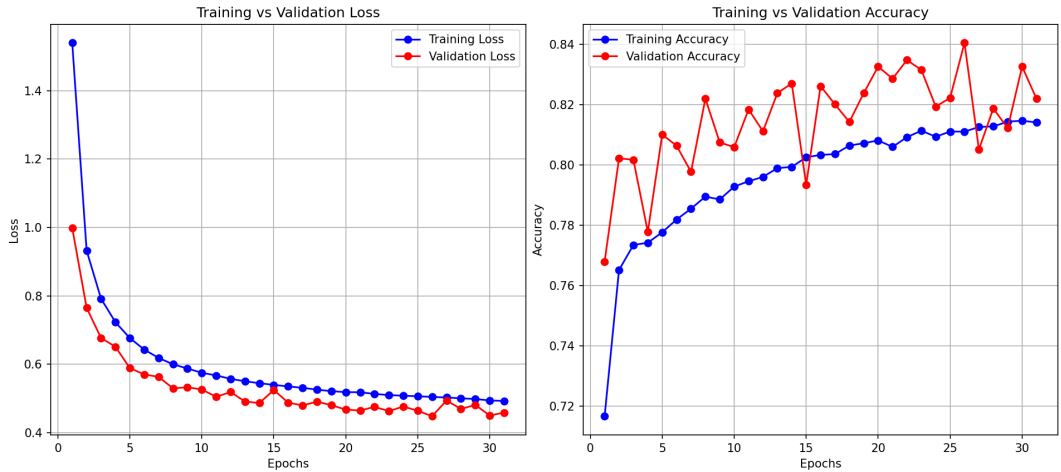

**Fig. 7** Loss and accuracy of training and validation of model 4

The 5-Fold Cross-Validation results demonstrate a mean accuracy of 79.8%, with notable inter-fold variance, providing validation that the model now operates without prior subject exposure (Kohavi 1995). Individual folds ranged from approximately 70% to 83%, suggesting that some subjects or data segments contain higher noise levels, but the model remains reliable on average.

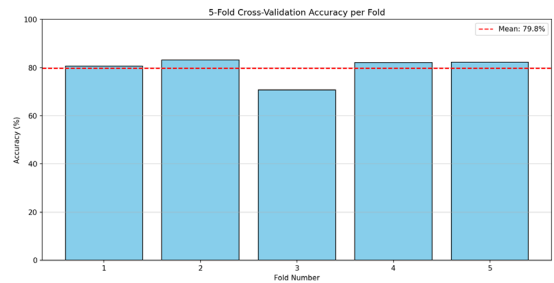

**Fig. 8** Cross-validation result of model 4

The Confusion Matrix reveals a significant class imbalance, with "Show warning" being the dominant class. The model correctly identified 3,689 instances of "Show warning," demonstrating high sensitivity for this critical category. There is a noted overlap between "Run as usual" and "Show warning," with 327 instances of "Run as usual" being misclassified. This suggests these two physiological states share similar feature signatures in the current dataset. Despite the confusion in specific predictions, the ROC curves indicate excellent separability between classes.

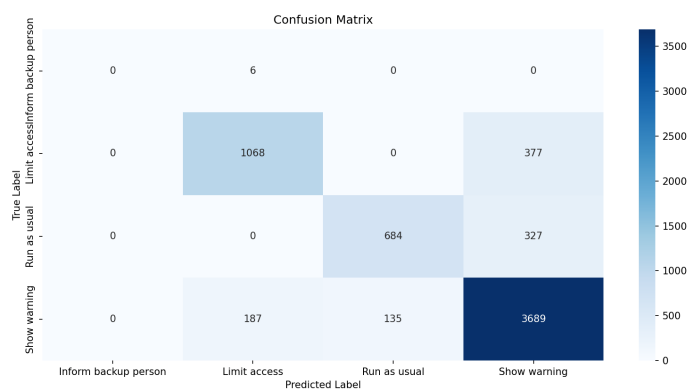

**Fig. 9** Confusion Matrix of model 4

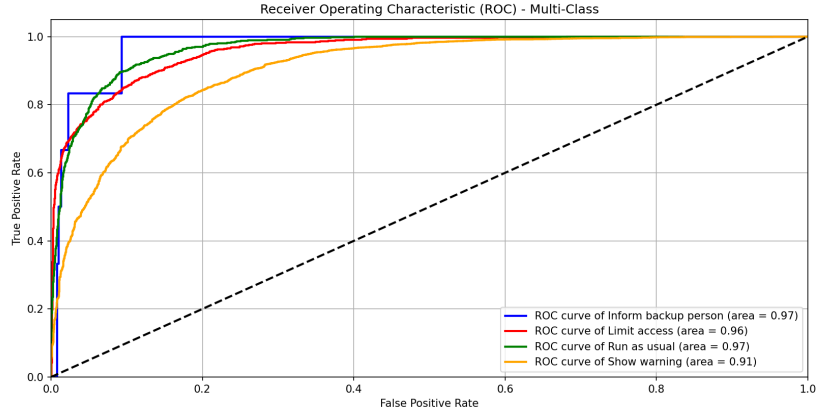

**Fig. 10** ROC curve of model 4

While the foundational research by Anicai and Shakir, using the "Health-spa" dataset, aimed to establish context-aware health monitoring by classifying environmental risks to cardiac health and estimating physiological indicators from ambient conditions, the present study shifts focus to human performance prediction, specifically targeting binary classification of user mistakes. In terms of performance, the baseline study achieved optimal results using a Random Forest algorithm, attaining 86.5% classification accuracy for cardiac risk levels and a Mean Absolute Error of 2.91 BPM for heart rate estimation. However, their reliance on traditional machine learning models, Random Forest and XGBoost, treats inputs as static features, potentially overlooking complex temporal dependencies inherent in physiological and environmental time-series data. In contrast, this research uses a hybrid CNN-LSTM architecture to process the same multimodal environmental and physiological data, aiming to capture a more comprehensive understanding of how both spatial patterns and temporal dynamics contribute to human error susceptibility. The CNN component extracts spatial features from sensor data, while the LSTM models long-range temporal dependencies that may indicate deteriorating performance states. Although this approach yields a slightly lower overall accuracy of 84%, this modest reduction is arguably justified by the increased model complexity and the fundamentally different prediction task, which involves forecasting discrete error events rather than continuous physiological states or categorical risk levels.

## 4 Model Deployment

To transition the developed CNN-LSTM model from a theoretical research environment to a functional real-time application, a modular Microservice Architecture was implemented. The model also serves as a container, separating the processor from the data-acquisition and visualization layers. This allows the system to remain stable and scalable by separating computationally intensive inference engine components. The deployment framework utilizes HTTP protocols and consists of three components: the Inference Server, the Application Gateway, and the Sensor Node.

### 4.1 System Architecture

The system architecture was designed to separate concerns between data handling and model execution, a strategy essential for preventing the heavy computational load of the LSTM network from freezing the user interface (Yin et al. 2024). The architecture is defined as follows:

1. The Inference Engine (Port 5000): A dedicated Flask API acting as the system's "Brain." It hosts the trained *Keras* model and handles all mathematical operations.

2. The Application Gateway (Port 8080): A secondary server acting as the "Traffic Controller." It manages data buffering, state management, and serves the web-based dashboard.

3. The IoT Sensor Node: A peripheral data source that streams physiological and environmental metrics to the Gateway.

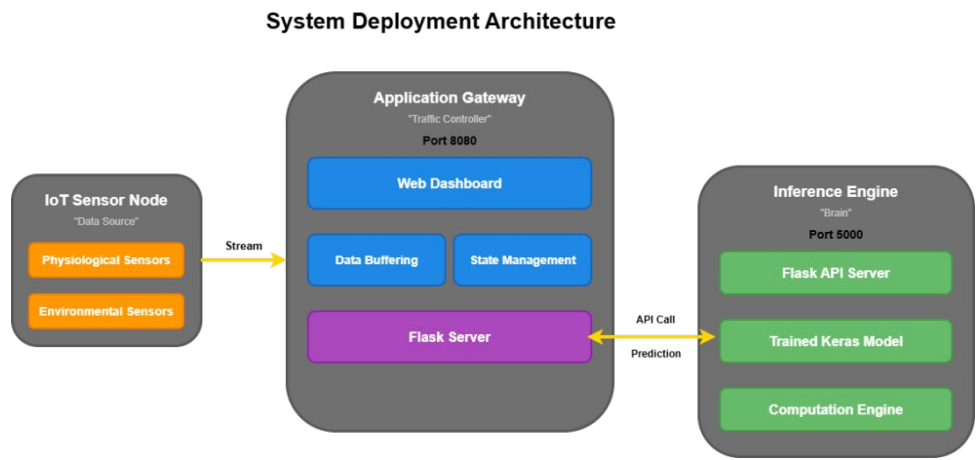

**Fig. 11** Model Deployment Architecture

## 4.2 The Inference Engine

The core analytical component is the Inference Engine, deployed using the Flask web framework. To optimize for latency, the system utilizes a **"Warm-Start"** initialization strategy. Upon server startup, the pre-trained *Risk_dectector.keras* model weights and *LabelEncoder* classes are loaded into memory immediately, rather than reloading them for each individual prediction request.

The engine exposes a RESTful endpoint that accepts raw sensor data batches and executes the following preprocessing pipeline:

1. Dynamic Standardisation: Incoming data is normalized using a local Z-score standardization method to account for inter-subject variability in real-time.

2. Tensor Reshaping: The 2D tabular data is transformed into the 3D tensor format (Samples, TimeSteps, Features) required by the LSTM layers.

3. Inference: The model computes the probability distribution over stress states, returning the predicted class and a confidence score to the Gateway.

## 4.3 The Application Gateway

To bridge the temporal disparity between the continuous sensor stream and the model's requirement for a complete session history, an Application Gateway was implemented on Port 8080. This middleware component functions as a stateful aggregator, designed to solve the issue where raw real-time data lacks the historical context necessary for accurate scaling.

In terms of functionality, the Gateway operates as a buffer to incoming telemetry into the memory until the predefined session threshold is reached. This means that instead of sending telemetry to the AI instantly, the Gateway is used as buffering. Maintaining this silence for the duration can prevent the system from predicting too soon or too quickly based on still incomplete data. The Gateway sends the complete dataset to the Inference Engine once the session is fully realized. This ensures that the user receives only a single definitive verdict, taking into account all the context of the event.

### 4.4 The IoT Sensor Node

To validate the system architecture prior to physical hardware deployment, a virtual sensor node was developed to simulate the data ingestion pipeline. This component utilizes a Python-based streamer that transmits physiological and environmental metrics to the application gateway at a precise frequency of 1Hz, mimicking the data throughput of a real-world IoT device.

Although the data transmission is simulated, the data itself is structured to strictly adhere to biological plausibility. The simulator utilizes synthetic augmentation strategies to ensure the input stream reflects the noisy, non-linear nature of human stress reactions. This rigorous simulation validates that the inference pipeline remains robust even when processing imperfect, real-world signal patterns.

### 4.5 Testing

To test and demonstrate the deployed CNN-LSTM model, users must first initiate the two-server architecture on their local machine by running the Flask applications via Python. The *Inference Engine* is launched by executing python App.py, which starts the model server on Port 5000 and loads the pre-trained *Risk_detector.keras* model into memory. Subsequently, the *Application Gateway* is deployed by running python Dashboard.py, which activates the dashboard server on Port 8080. Once both servers are running, users can access the real-time monitoring dashboard by navigating to http://localhost:8080 in any web browser, where they will find the visualization interface displaying prediction results and sensor metrics. To begin data streaming and generate predictions, the *Virtual Sensor Node* is activated by executing python sensor.py in a separate terminal window, which initiates the transmission of simulated physiological and environmental data at 1Hz intervals to the *Gateway*. As the sensor streams data, users can observe real-time updates on the dashboard, including incoming sensor readings, data buffering progress, and ultimately the model's binary prediction along with confidence scores once sufficient session data has been accumulated. This local deployment setup allows for comprehensive system testing without requiring physical IoT hardware, while maintaining complete functional parity with a production-ready implementation.

## 5 Limitations and Future Directions

One fundamental problem with the model's training performance is that there exists a severe class imbalance. This is mainly due to the training data origin; the *Health-Spa* data was originally developed for general environmental health monitoring purposes instead of specific risk stratification. As a result, the data distribution shows long spells of homeostasis interrupted by only brief, specific environmental variations, resulting in a dataset where normal conditions vastly exceed critical events. This imbalance is exacerbated by the custom labelling heuristic applied in this study. Because the risk scores were obtained mathematically from a strict set of environmental and physiological thresholds, criteria for the "Inform a backup person" class were met relatively infrequently compared to baseline conditions. Because fewer minority-class examples are available than necessary, the model can end up tilting its weights toward the majority. This may not be able to see rare but grave high-stress anomalies without recourse to synthetic oversampling techniques.

The system's reliance on dynamic standardization presents a duality of both advantage and limitation. During the training phase, this approach is a significant asset; the availability of moderate-length data sequences allows the scaler to effectively identify and neutralize individual physiological baselines, thereby enabling the model to generalize across diverse subjects without memorizing specific biological signatures. However, in a real-time IoT deployment scenario, this strength risks becoming a vulnerability due to data scarcity. When the system receives only a short, homogeneous stream of high-stress data, the scaler may interpret the elevated signals as the user's new normal baseline. However, in the final deployment, this limitation is effectively mitigated by enforcing a dataset that captures both a relaxed initialization phase and the subsequent stress event. By integrating data from both normal and high-risk conditions into a single scaling context, the system provides the model with a comprehensive physiological picture, ensuring that stress deviations are correctly identified against a valid baseline rather than being normalized away.

Furthermore, a major limitation of this study is the lack of construct validity for the ground-truth labels, which are based on a theoretical scoring heuristic rather than empirical observation. The composite risk scores were calculated using a fixed-weighting approach, in line with the literature in psychophysiology. Although this method establishes a scientifically grounded baseline, it inherently assumes a uniform response profile across all subjects, thereby omitting inter-subject variability in environmental tolerance and stress resilience. Due to the use of a secondary dataset in the current analysis, without concurrent subjective self-reports or biochemical validation, it was not feasible to empirically calibrate these weights to reflect participants' actual experiences. As a result, the model predicts a theoretical risk state based on parameters from the literature, which may, in some cases, differ from a user's actual subjective state of distress or cognitive load.

To overcome the inherent challenges of intersubject variability, future iterations of the system should move from generic, subject-independent modelling to subject-specific adaptation. Implementing a 24-hour calibration phase upon initial deployment can address the aforementioned problem. During this period, the user would wear the device for a full day to capture a comprehensive longitudinal profile of their unique physiological baseline, covering sleep, low-activity work, and high-intensity movement. This data would facilitate transfer learning, in which the pre-trained CNN-LSTM weights are frozen, and the final decision layers are fine-tuned on the user's specific data. This approach would effectively transform the universal detector into a personalised detector, establishing accurate detection while significantly reducing false positives caused by natural physiological variations.

The current model relies on the *Health-Spa dataset*, which primarily simulates stress through environmental factors in a controlled setting. Although this dataset works well to confirm the construct of the architecture, it could also fail to convey the nuance of psychosocial or cognitive stress in everyday life. Further research is to conduct a broader data collection to develop a proprietary dataset of greater ecological validity. This project would involve monitoring participants in less-constrained, real-world settings over longer periods. By incorporating this broad spectrum of stress response-generating features, from mental tasks, social interaction, to physical effort, we get more features to train the AI. Such variation in diversity will improve the model's generalization performance in separating the actual cognitive burden from physiological excitation and arousal caused by events such as non-threatening physical activity from a real psychological stressor.

**6 Literature Review**

Recent literature has increasingly highlighted the detrimental impact of acute stress on decision-making processes. Wemm and Wukfert demonstrated that under stressful conditions, cognitive processing becomes biased toward immediate rewards. At the same time, sensitivity to potential punishment is impaired (Wemm and Wulfert 2017). This suggests a tendency to prioritize speed and completion, such as rushing to finish an exam, while failing to consider or correct potential errors.

Contemporary stress assessment increasingly prioritizes physiological biomarkers, with heart rate, blood pressure, and GSR serving as the most significant indicators of the body's internal state. Gedam and Paul corroborate this in their comprehensive review, demonstrating that capturing these specific signals through wearable technology and processing them with machine learning algorithms yields high diagnostic accuracy for mental stress (Gedam and Paul 2021). These findings underscore the pivotal role of Heart Rate and GSR as objective metrics, as they directly capture the autonomic nervous system's reactivity in real-time, offering immediate insights that subjective self-reports often miss. However, the researchers identify that signal noise and motion artifacts remain significant limitations in single-sensor setups, ultimately advocating for a multimodal sensor fusion approach—combining multiple data streams simultaneously—as the necessary solution to ensure robust and reliable detection

Beyond internal physiological states, the external physical environment exerts a quantifiable pressure on cognitive bandwidth, primarily through the mechanisms of thermal comfort, acoustic interference, and lighting spectrum. Seppänen et al established a distinct physiological threshold for productivity, identifying that cognitive performance degrades by approximately 2% for every degree Celsius increase above an optimal 22°C (Seppänen et al. 2006). This thermal burden is often compounded by auditory distractions; Jahncke et al. demonstrated that 'irrelevant speech' in open environments forces involuntary semantic processing, resulting in significantly reduced working memory and increased fatigue (Jahncke et al. 2011). Furthermore, the spectral quality of the visual environment plays a regulatory role in alertness. Chellappa et al. found that exposure to blue-enriched, high-correlated colour temperature (6500K) lighting actively suppresses melatonin and enhances sustained attention compared to warm lighting (Chellappa et al. 2011). Collectively, these studies suggest that cognitive fatigue is not merely a product of task difficulty, but often a symptom of sub-optimal environmental regulation.

Crucially, the degradation of cognitive performance in sub-optimal environments is not merely a product of psychological distraction but is directly mediated by physiological arousal mechanisms. The 'silent' cost of the environment was demonstrated in a landmark study on low-intensity office noise;. At the same time, participants did not report subjective feelings of stress; their urinalysis revealed significantly elevated epinephrine levels, indicating that the body was mounting a neuroendocrine stress response despite a lack of conscious awareness (Evans and Johnson 2000). This finding is paralleled in thermal research observing that deviations from thermal comfort precipitated a significant reduction in Heart Rate Variability (HRV)—a key marker of autonomic nervous system balance (Liu et al. 2013). Together, these studies confirm that environmental stressors like noise and heat bypass conscious appraisal and directly trigger the sympathetic nervous system, inducing the exact state of physiological dysregulation identified as the precursor to risky and maladaptive decision-making (Wemm and Wulfert, 2017)

While traditional cybersecurity frameworks prioritize technical defences such as firewalls and intrusion detection systems, recent literature identifies the human operator as the most vulnerable component of the security infrastructure. Khadka and Ullah argue that human behaviour, including stress, cognitive overload, and fatigue, significantly contributes to insider threat incidents (Khadka, K., and Ullah 2025). Their review highlights that static security policies fail to account for the dynamic nature of human performance, which fluctuates based on physiological state.

Recent experimental studies have empirically validated this theoretical link between physiological state and security vulnerability. Wiemken et al. investigated the impact of emotional manipulation on phishing susceptibility in a simulated email environment, recording multimodal signals including facial expressions and EDA (Wiemken et al. 2025). Their statistical analysis revealed a significant correlation between elevated physiological stress markers and decision-making errors, such as failing to flag malicious emails or impulsively replying to phishing attacks. Additionally, another study found that emotionally charged triggers, such as fear and urgency, successfully bypassed users' cognitive defences, leading to a higher rate of victimization despite security awareness (Lain et al 2021). These findings confirm that cybersecurity is not merely a technical challenge but a psychological and physiological one, necessitating adaptive systems that can monitor the user's real-time capacity to withstand emotional manipulation and maintain security protocols.

To objectify the measurement of human stress, researchers have increasingly turned to deep learning architectures capable of modelling complex biological signals without manual feature engineering. Mane and Shinde used a hybrid CNN-LSTM model, StressNet, to predict stress from EEG (Mane and Shinde 2023). In this architecture, a CNN is employed to extract spatial features, such as morphological patterns in signal spectrograms or 2D projections, effectively acting as a noise filter and a high-level feature extractor. These spatial features are then fed into LSTM networks, which excel at modelling temporal dynamics and recognizing past states. Their model achieved an accuracy of 97.8% on benchmark datasets. This validates the choice of a hybrid architecture for detecting dynamic stress events in continuous data streams.

Despite the advancements in both affective computing and insider threat mitigation, a distinct gap remains at their intersection. The existing literature is largely bifurcated: cybersecurity studies focus on the behavioural and psychological drivers of insider threats without access to real-time physiological ground truth, while biomedical studies develop highly accurate stress-detection algorithms without a specific application in the cybersecurity field. Furthermore, a critical review of current stress detection methodologies reveals significant limitations, including subject-dependent training and reliance on single-modality sensors, leading to poor generalisation to unseen users and increased sensitivity to noise. Crucially, most approaches ignore the environmental context, thereby reducing the explainability of the prediction. This project addresses these limitations by developing a CNN-LSTM designed specifically for the cybersecurity domain, leveraging a deeper integration of environmental context to enhance predictive reliability and reduce false positives in real-time threat mitigation.

## 7 Conclusion

The paper proposes a security framework that integrates a hybrid Convolutional Neural Network–Long Short-Term Memory (CNN–LSTM) model to analyze biometric and environmental data for context-aware security decisions to reduce cybersecurity incidents driven by human behaviour. This conceptual security framework integrates soft biometric data with environmental sensing to estimate users' emotional and cognitive states in real time. When elevated risk is detected, the framework can initiate protective actions, such as issuing warnings, providing additional instructions, enabling backup support, or temporarily restricting access. This framework aims to reduce non-intentional insider threats while maintaining usability. The model uses the Health-Spa dataset, which primarily simulates stress through environmental factors in a controlled setting. This approach achieves 84% accuracy, demonstrating that the model can reliably detect conditions that lead to elevated cyber risk. By enabling continuous monitoring and adaptive safeguards, the framework supports proactive interventions that reduce the likelihood of human-driven cyber incidents.

**Availability of data and materials**

The dataset analyzed during the current study is publicly available in the SpringerNature Figshare repository: Anicai C, Shakir MZ (2023) A Multimodal Dataset of Cardiac, Electrodermal, and Environmental Signals. https://doi.org/10.1038/s41597-025-05051-3

**Competing interests**

The authors declare that they have no competing interests.

**Funding**

This research was supported by Western Sydney University Summer Research Scholarship Program SCDMS.

**Authors' contributions**

Duy designed the methodology, developed the CNN-LSTM model, tested different models, implemented the deployment architecture, and wrote the manuscript. Farnaz supervised the research, provided guidance on methodology and analysis, and revised the manuscript. All authors provided support tools and guided through model development phase, read and approved the final manuscript.

**Acknowledgements**

The authors would like to thank Cezar Anicai and Muhammad Zeeshan Shakir for making their Health-Spa dataset publicly available, which was instrumental to this research.